# Multispecies bird sound recognition using a fully convolutional neural network

María Teresa García-Ordás[1], Sergio Rubio-Martín[2], José Alberto Benítez-Andrades[2*†], Hector Alaiz-Moretón[1†] and Isaías García-Rodríguez[1†]

[1]SECOMUCI Research Group, Escuela de Ingenierías Industrial e Informática, Universidad de León, Campus of Vegazana s/n, León, 24071, León, Spain.
[2*]SALBIS Research Group, Department of Electric, Systems and Automatics Engineering, Universidad de León, Campus of Vegazana s/n, León, 24071, León, Spain.

*Corresponding author(s). E-mail(s): jbena@unileon.es;
Contributing authors: mgaro@unileon.es;
srubim00@estudiantes.unileon.es; halam@unileon.es;
igarr@unileon.es;
†These authors contributed equally to this work.

**Abstract**

This study proposes a method based on fully convolutional neural networks (FCNs) to identify migratory birds from their songs, with the objective of recognizing which birds pass through certain areas and at what time. To determine the best FCN architecture, extensive experimentation was conducted through a grid search, exploring the optimal depth, width, and activation function of the network. The results showed that the optimal number of filters is 400 in the widest layer, with 4 convolutional blocks with maxpooling and an adaptive activation function. The proposed FCN offers a significant advantage over other techniques, as it can recognize the sound of a bird in audio of any length with an accuracy greater than 85%. Furthermore, due to its architecture, the network can detect more than one species from audio and can carry out near-real-time sound recognition. Additionally, the proposed method is lightweight, making it ideal for deployment and use in IoT devices. The study also presents a comparative analysis of the proposed method against other techniques, demonstrating an improvement of over 67% in the best-case scenario. These findings contribute to advancing the field of bird sound recognition and provide valuable insights into the practical application of FCNs in real-world scenarios.

# 1 Introduction

A large number of landbirds often undertake long-distance migrations. There are a number of common patterns among many of them, e.g., flying north to breed in summer in colder or arctic areas and returning to warmer southern regions in winter [1]. Migration, in terms of its physiology, involves endogenous processes that are triggered by external stimuli received by the central nervous system.

There are issues following the global onset of climate change. Some studies are observing chronological changes in the migratory processes of birds, a reduction in populations and even changes in the breeding season [2].

The most recent forecasts released by BirdLife International and the National Audubon Society show that 10% of the world's bird species may be in danger in the next century due to global warming [3].

At higher temperatures, flowers and plants bloom earlier, and insects also develop earlier; this means that birds have to adapt and arrive earlier in Africa to be able to eat, because they must adjust the reproduction period to coincide with that of maximum available food [4, 5]. Many birds fail to do so, and this is creating conservation problems for certain species. Faced with this new situation, migratory birds can do three things to adapt to the present situation: fly faster, leave sooner or fly a shorter distance, that is, shorten migration.

The management of natural resources is key, and the conservation of birds is also crucial for the future of humanity [6]. It is important to promote the protection of migratory birds because of the vital economic and environmental benefits that they maintain. A change in the migration of birds could become a serious problem that greatly affects humans [7].

For this reason, studies related to the migratory processes of birds have been carried out using different techniques. Banding is the oldest known technique. Other common techniques in the study of migrations include satellite tracking, colour marking, analysis of stable isotopes of hydrogen (or strontium) and the use of radar [8].

One method for determining migratory intensity makes use of microphones with a specific orientation, upwards, with which contact calls between flying flocks are recorded throughout the night. After these recordings are made, time, frequency and species measurement analyses are carried out in specialised laboratories [9].

This article proposes a method to detect the different migratory birds that pass through our country, more concretely through Villafáfila, and identify the sound of their song. In this way, it is possible to determine which species are being affected by climate change and may be in danger of extinction; they may be unable to breed because they cannot migrate to find the necessary food for themselves or their young. The disappearance of certain species can have direct consequences for life on our planet, and therefore its study is crucial.

The proposed method is based on a fully convolutional neural network architecture and offers several advantages over state-of-the-art methods:

- **Flexible audio processing**: The proposed method can process audio of any length and identify migratory bird species using fully convolutional neural networks, eliminating the need to preprocess the bird sounds to a fixed length. This feature offers significant advantages, as it makes the method highly adaptable to different environments and scenarios.
- **Near-real-time detection**: The proposed method does not require sounds to have a fixed length, enabling it to perform near-real-time detection by processing small fragments of audio. The model can accurately determine the bird species even in the presence of high-frequency sounds. This feature is especially useful in situations where real-time monitoring is critical, such as wildlife conservation and management.
- **Robustness in complex backgrounds**: The proposed method is highly robust and can accurately detect bird species even in the presence of background noise or silence. The model achieves this by removing these parts for classification through input size flexibility or creating a non-bird class for these parts. This feature enhances the method's accuracy and applicability in real-world environments.
- **Lightweight and efficient**: The proposed method is highly efficient and does not require a significant amount of computational power. With fewer than 500,000 trained parameters, the method avoids the use of fully connected layers in the neural network, making it a lightweight and efficient solution for migratory bird detection.
- **IoT deployment**: The proposed method is suitable for deployment on IoT devices. This feature enables remote monitoring and real-time bird species detection in remote areas, further extending the applicability of the method.

Overall, these advantages make the proposed method highly effective and versatile, and it is suitable for a wide range of applications, from wildlife conservation to environmental monitoring.

A graphical abstract with a summary of the work carried out is shown in Figure 1.

The rest of the paper is organized as follows: Section 2 summarizes related work connected with this field of research. In Section 3, the dataset and the methods used are described. The experiments and results are shown and discussed in Section 4, and finally, conclusions are stated in Section 5.

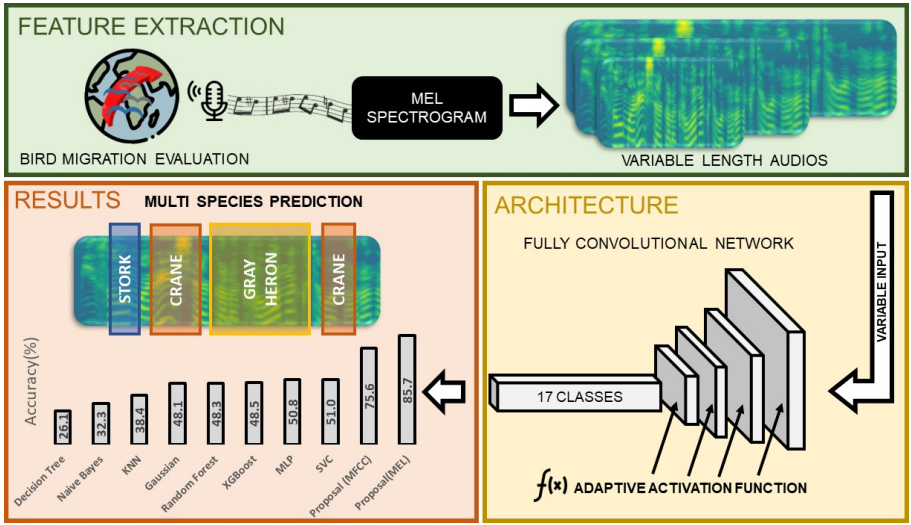

**Fig. 1** Graphical abstract of the work carried out

# 2 Related work

Much work has been performed to detect bird species based on their songs. The best-known method is probably BirdNET [10]. In this work, Kahl et al. developed a deep neural network derived from ResNet that has the ability to identify a large number of species, 984, of birds from Europe and North America. This model is composed of 157 layers with approximately 27 million parameters. The training of this model involves very advanced data preprocessing and the use of augmentation and blending techniques.

Deep learning was also used in [11], where the authors proposed a new model called AMResNet for automatic recognition of bird sounds. Their proposed model fused the ResNet model with different numbers of attention layers in various channels and spaces to improve bird sound classification. Moreover, Log-mel, MFCC, chroma, spectral contrast and Tonnetz features were combined into three different feature sets to improve the final classification performance.

To reduce the computational burden of deep learning methods, in [12], the authors used traditional machine learning techniques. A model called Iterative ReliefF (IRF), based on ReliefF, was used with a number of improvements. It is a feature-selective model since, during its operation, IRF automatically selects the features that contain the most information. Subsequently, these features are used on different classification models such as k-nearest neighbours (KNN), support vector machines (SVMs), bagged trees (BTs) and linear discriminants (LDs), to calculate the results of the variable classifiers.

In the same way, Hsu et al. [13], proposed the local wavelet acoustic pattern (LWAP), which is a new sound descriptor created with the aim of characterising the correlation of DWT coefficients across different subbands and, thanks to this characterisation, to recognise the songs of different birds. In this study, they used techniques to reduce the dimensionality of the features for classification, such as linear discriminant analysis (LDA) and principal component analysis (PCA).

A different but closely related problem was formulated by Briggs et al. [14]: the problem they solved in their research was to classify a set of species from an audio recording. To do this, they used a multi-instance multilabel (MIML) framework for machine learning, and, in turn, proposed a MIML bag generator for audio. This approach resulted in an algorithm capable of transforming an input audio signal into a bag representation of instances suitable for use with MIML classifiers. This is because recordings collected in the field often include a multitude of birds of different species vocalising at the same time.

Although they incur a large computational load, methods based on deep learning are most often used to solve current problems, including medical problems [15–17], quality control problems [18], and problems in other areas [19, 20], and they can also address the problem of bird song classification.

Regarding the problem at hand, deep convolutional neural networks (CNNs) were used in [21] to identify bird sounds and achieved good performance, but recognition was only employed to distinguish woodpeckers. The recognition of a single species does not approximate the problem we want to solve, but there are other works where several species are differentiated.

For example, CNNs were also used in [22] with the aim of detecting 14 forest-adapted birds and mammals automatically using methods that classify spectrogram images generated from short audio clips. In addition, in [23], the authors developed an efficient data processing pipeline using a deep convolutional neural network (CNN) to automate the detection of owl vocalizations in spectrograms generated from unprocessed field recordings. Additionally, Zhang et al. [24] differentiated between four different species of bird sounds using a continuous frame sequence and a spectrogram-frame linear network (SFLN). To create the continuous frame sequence that serves as the standard input for the SFLN, the authors employed a sliding window algorithm with a short frame length to differentiate the mel-spectrogram of bird sounds. In the linear layer, a vertical 3D filter moves linearly along the continuous frame and covers its entire frequency range. The weight of the filter is initialized as a Gaussian distribution to reduce high- and low-frequency noise, which allows for the extraction of the long- and short-term features of the bird sound's continuous frame. In [25], the authors combined acoustic features, visual features, and deep learning to solve the problem of bird sound classification. Finally, Okan and his group [26] proposed a system that can record data, then perform preliminary on-board signal preprocessing, extract features, and classify and store data. This system is composed of a Texas Instruments Tiva C microcontroller, a storage unit, and a microphone and it was also used for the bird sound classification problem.

After reviewing the most recent literature, we chose fully convolutional neural networks (FCNs) for the audio tasks. This type of network has been widely used in recent years for different purposes. An FCN tries to learn representations and make decisions based on local spatial input. Appending a fully connected layer enables the network to learn using global information, where the spatial arrangement of the input falls away and need not apply. In [27], a sentiment analysis method was proposed that is capable of accepting audio of any length, without being fixed a priori. This is the same goal that we pursue in this work. Fully convolutional networks were first developed to perform emotion classification on three known datasets (RAVDESS, EMODB, and TESS) and second to allow near-real-time sentiment analysis in order to determine the evolution of a conversation, which is of interest for many companies such as banks, call centres, and even hospitals. In [28], the authors proposed two FCN-based architectures for sound event detection. The advantages of FCNs are as follows: other architectures are confined to the duration of time dependencies, resulting in a failure to model sound events with long durations. However, the use of FCNs allows temporal dependencies to be captured, and FCNs have demonstrated a strong ability to ensure high time resolution and obtain longer temporal dependencies with unchanged filter size and network depth.

FCNs are also used in the medicine field, among other fields. In [29], a new fully convolutional modular deep neural network for brain tumour diagnosis based on nuclear magnetic resonance (MRI) was designed. The proposed network consists of four modules: feature extraction (FE), residual strip pooling attention (RSPA), atrous spatial pyramid pooling (ASPP), and a classification module. First, the FE module extracts the discriminatory features of brain tumours using multiple residual convolutional blocks, and then the RSPA module strengthens prominent tumour regions that are relevant for brain tumour classification. The ASPP module captures features at various scales that contain useful contextual information. Finally, a classification module is adopted that uses convolutional layers with adjusted stride values to classify the extracted multiscale features. The combination of these modules helps extract local and contextual information that is appropriate for the classification of brain tumours. In this case, the use of FCNs can solve the problems of large intraclass variation and small dataset sizes in medical image classification. Additionally, fully convolutional networks have been successfully used in various fields, including liver and tumour segmentation [30], melanoma segmentation [31], and retinal vessel segmentation [32]. FCNs have become a popular and effective tool in these fields.

In summary, the recent literature shows the widespread use of fully convolutional neural networks (FCNs) in various applications, including sound event detection, sentiment analysis, and medical image classification. FCNs have shown great promise in medical imaging applications, where small dataset sizes and large intra-class variations are common challenges. Furthermore, FCNs can capture temporal dependencies and demonstrate a strong ability to ensure high time resolution, making them well-suited for the analysis of audio data.

A summary of the related work is shown in Table .

**Table 1** Summary of works on bird species detection using different methods

| Authors | Year | Method | Features |
|---|---|---|---|
| Xiao et al. [11] | 2022 | Deep learning | Log-mel, MFCC, chroma... |
| Kahl et al. [10] | 2021 | Deep learning | Advanced preprocessing |
| Tuncer et al. [12] | 2021 | Traditional ML | ReliefF |
| Ruff et al. [22] | 2021 | Deep learning | Spectrogram |
| Juliette et al. [21] | 2020 | Deep learning | Spectrogram |
| Ruff et al. [23] | 2019 | Deep learning | Spectrogram |
| Zhang et al. [24] | 2019 | Deep learning | mel-spectrogram |
| Xie et al. [25] | 2019 | Deep learning | Acoustic and visual features |
| Kucuktopcu et al. [26] | 2019 | Embedded system | MFCC |
| Hsu et al. [13] | 2018 | Feature engineering | LWAP, LDA, PCA |
| Briggs et al. [14] | 2012 | MIML | Spectrogram |

# 3 Methods

## 3.1 Datasets

In this work, a study on the main migratory species that pass through Spain is carried out; we compiled their songs from the Xeno-canto database [1], which is a website dedicated to sharing bird sounds from all over the world, and we created our own dataset made up of 2699 samples and 17 different classes, as shown in Table 2.

**Table 2** Number of samples per species.

| Species | #Samples |
|---|---|
| Golondrina común | 201 |
| Avión común | 201 |
| Vencejo común | 201 |
| Aguila calzada | 44 |
| Cigüeña | 94 |
| Cigüeña negra | 25 |
| Grulla | 201 |
| Gaviota reidora | 201 |
| Flamenco común | 173 |
| Ánade real | 201 |
| Garza real | 201 |
| Ánsar común | 201 |
| Correlimos | 208 |
| Cernícalo primilla | 106 |
| Cuchara común | 131 |
| Aguilucho cenizo | 108 |
| Andarríos chico | 202 |

The audio files were trimmed to a maximum of 20 seconds to create a dataset with the following duration statistics: The mean length of an audio file was 15.82 seconds, with a standard deviation of 5.70. The smallest audio file was 0.74 seconds long, and 50% of the files were longer than 20 seconds and were therefore trimmed. The first quartile (Q1) of the duration distribution was 11.48 seconds.

---

[1] https://www.xeno-canto.org/

It is important to note that a preprocessing step was performed to remove all periods of silence from the audio files, so the final duration of a file corresponds to the time during which a bird was singing.

## 3.2 Description methods

In preprocessing the audio files before they were fed into the neural network, the mel spectrogram and the mel frequency cepstral coefficients (MFCCs) were used.

### 3.2.1 Mel Frequency Cepstral Coefficients (MFCCs)

The mel frequency cepstral coefficient (MFCC) descriptor [33] is one of the most commonly used voice parametrization techniques in terms of verification. The main goal of this transformation is to obtain a compact, robust and suitable representation so that after this description process, a statistical model of the speaker can be obtained with a high degree of precision.

The audio description process comprises six steps, represented in Figure 2.

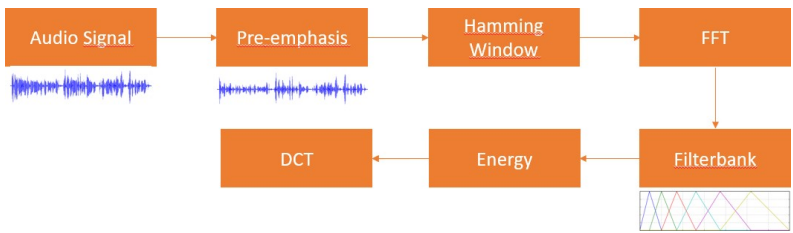

**Fig. 2** Mel frequency cepstral coefficient extraction steps

- Pre-emphasis: A filter (see equation 1) is applied to the audio signal to compensate for the attenuation that occurs in the sound production mechanism. This step is useful to emphasize higher frequencies.

$$H(z) = 1 - bz^{-1} \qquad (1)$$

The value of *b* should be between 0.4 and 1 to control the slope of the filter.
- Hamming window: Audio signals are nonstationary, and this can make it inconvenient to analyze them. In a short period of time (on the order of ms) the signal is quasi-stationary. Therefore, to facilitate analysis, windowing is usually performed, which involves dividing the signal into segments of a few ms. Generally, Hamming windows with a size of 20 ms are used. Also it is useful to carry out the windowing with blocks of samples that overlap each other so that no information is lost in the transition between windows. Generally, the overlap is performed with an offset between windows of 10 ms, yielding MFCC coefficients every 10 ms.

- Fast Fourier Transform: After the windowing, the FFT of size N of the windowing signal is calculated by equation 2:

$$X_k = \sum_{n=0}^{N-1} x_n e^{\frac{-j2\pi nk}{N}} \quad (2)$$

where $N$ is the number of points used to compute the DFT and $k$ is between 0 and $N \pm$.
- Filterbank: The signal obtained in the previous step is multiplied by a bank of triangular filters of unit area. The triangles are spaced according to the mel frequency scale.
- Energy: After the speech signal spectrum is multiplied by the filterbank, the corresponding energy is calculated in each of the filters following equation 3:

$$E = \sum_{k=0}^{N-1} X_k^2 H_{mk} \quad (3)$$

where $m$ is between 1 and the number of filters. After obtaining the energy, we must calculate the logarithm and then pass both to the domain of logarithmic spectral power. The downside of work in this domain is that the spectra of the filters in the adjacent bands show a high degree of correlation, giving rise to spectral coefficients that are statistically highly dependent on each other. To address this drawback, the discrete cosine transform is applied.
- Discrete Cosine Transform: The spectral coefficients are transformed to the frequency domain by converting them into cepstral coefficients (MFCCs) (see equation 4):

$$C_{mfcc}[m] = \sum_{k=}^{N-1} \log(E_k) \cos\left(m\left(k - \frac{1}{2}\right)\frac{\pi}{N}\right) \quad (4)$$

where $m$ is between 1 and the number of filters.

### 3.2.2 Mel spectrogram

The mel frequency scale is represented with equation 5.

$$mel(f) = 2595 \log_{10}\left(1 + \frac{f}{700}\right) \quad (5)$$

Mel spectrograms are spectrograms characterised by the fact that their frequencies are converted to the mel scale, which is the origin of the name mel spectrogram. In this type of spectrogram, the audio signal is mapped from the time domain to the frequency domain using the fast Fourier transform. To apply this technique, it must be performed on segments of the audio signal with overlapping windows, as explained in the previous section for MFCC extraction. The frequency is converted to a logarithmic scale, and the amplitude is converted to decibels. In this way, the spectrogram is obtained. Finally, the frequency is converted to the mel scale to obtain a spectrogram of this type.

Different audio descriptors have been taken into account, such as empirical mode decomposition (EMD) or EEMD, but have shown suboptimal results. The use of mel spectrogram EMD is justified in this study due to its effectiveness for capturing the tonal and timbral characteristics of bird vocalizations and its wide usage in studies of bioacoustics. The audio data used in this study are comprised of bird vocalizations, which are characterized by complex and varying frequencies over time. While EMD is a powerful technique for analysing non-stationary signals, it may not be the most appropriate approach for analysing the bird vocalizations in this study, as they are characterized by complex frequency patterns that cannot be easily decomposed into intrinsic mode functions (IMFs) using EMD. Moreover, EMD may introduce spurious IMFs or aliasing artifacts, which can lead to erroneous analysis of the bird vocalizations. In contrast, mel spectrograms have been shown to be effective for capturing the tonal and timbral characteristics of bird vocalizations, and have been widely used in studies of bioacoustics. Therefore, the use of mel spectrograms is considered a more appropriate approach for analysing the bird vocalizations in this study, as it is better suited to capturing the complex frequency patterns of the vocalizations while also avoiding potential issues with EMD. Due to the suboptimal performance observed with this method, the results obtained were not included in the paper.

### 3.3 Fully Convolutional Neural Network

In this paper, a fully convolutional neural network (FCN) is used to describe and classify bird sounds. The FCN architecture is similar to a convolutional neural network (CNN) architecture, but it is different in that all the learnable layers are convolutional layers. It is common in CNNs for the last layers to be fully connected layers. In the case of an FCN, these fully connected layers do not exist. Pooling layers can be part of the FCN because these types of layers do not have weights or parameters to be learned. They usually also include regularization layers such as dropout or batch-normalization layers.

The main reason we used this FCN architecture is that if the network has only convolutional layers and does not have any fully connected layers, input data matrices of any size can be received as input, since the only layers that require the size to be fixed are the fully connected layers. Therefore, when using the FCN, we can analyze any bird song sound without having to set the size of the audio.

Additionally, in the case of CNNs, there is a loss of spatial information, since in the fully connected layers, all the outputs of one layer are connected to all the inputs of the next layer.

In our case, the key point of FCNs for enabling variable input sizes lies in the use of a global pooling layer at the end.

Two kinds of global pooling layers can be used: a global max pooling layer and a global average pooling layer. These kinds of layers resize the dimensions of the input, (height, weight, nfilters), to (1,1,nfilters). In the case of a global max pooling layer, these values are obtained by taking the maximum value in the height and width dimensions for every filter, and in the case of global average pooling, the average value is taken instead of the maximum. Using this layer, we can determine the output size of our neural network and match it with the number of classes to train a classification problem regardless of the input size. A general representation of a vanilla fully convolutional neural network is shown in Figure 3.

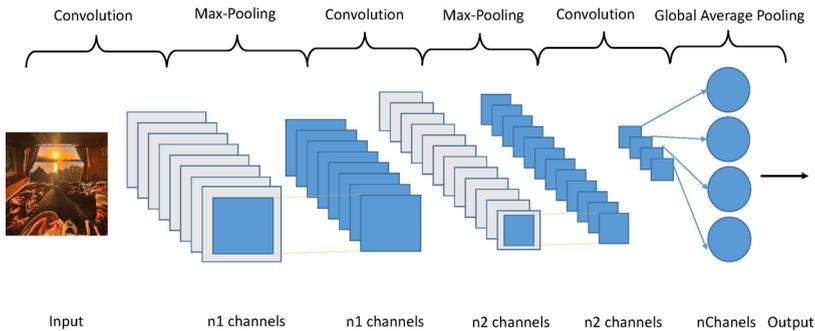

**Fig. 3** Vanilla FCN representation

# 4 Experiments and results

## 4.1 Experimental Setup

### 4.1.1 Data preprocessing

The audio files were described in two ways: using the mel frequency cepstral coefficients (MFCCs) and using the mel spectrogram. In both cases, the LibROSA [34] Python library is used to load the audio files. A sample rate of 44100 Hz was chosen because it is considered the highest usable rate supported by most devices. [35]. After that, the audio is trimmed to leave only the segments in which bird songs are heard, eliminating silences from the audio. For training, all the audio files were modified to the same size by filling the end of each file with silence up to the length defined by the longest audio in the dataset. No further preprocessing was carried out related to the length of the testing audio files, since the use of the FCN allows the processing of audio files of any length, and in turn, it is possible to carry out almost real-time recognition of the sounds of migratory birds that are heard in a certain audio file, that is, to be able to determine the bird that is singing with a very high frequency throughout the entire audio.

### 4.1.2 Network Architecture

The architecture of the final neural network was selected after an exhaustive grid search. We explored different combinations of hyperparameters such as the number of layers, the size of the filters, and the activation functions used in each layer. Specifically, we tested 3, 4, and 6 convolutional layers with filter sizes of 100, 250, and 400 and activation functions including ReLU, TanH, and the adaptive activation function [36]. We evaluated all of them using as input the MFCC and mel spectrogram features. After evaluating all the possible combinations, we found that the best network consisted of 4 convolutional layers with 100, 400, 100, and 17 neurons, using the adaptive activation function in all layers except the last one, which used softmax. The size of the kernels is 3 x 3 in all cases except in the last convolutional layer, which has a kernel size of 1. The "same padding" configuration was used to allow processing of smaller audio chunks without facing the problem of data size reduction in each convolutional layer. After each convolutional layer except the final one, maxpooling was applied, and a global average pooling layer was added in the final layer to reduce the data and prepare the model for the output classification layer using a softmax activation function. To avoid overfitting, a dropout of 0.4 was applied in the final layer. With this architecture, the output size of the neural network is matched to the number of classes, enabling the training of a classification problem regardless of the input size.

### 4.1.3 Training setup

To ensure that this network would be generalizable, 10-fold Monte Carlo cross-validation was performed. An 80-20 data distribution was used for training and testing, respectively, and precision was used as a performance metric to evaluate the different approaches.

In the experiments performed in this research, cross-entropy was chosen as the loss function and Adam as the optimiser. Five hundred epochs were carried out with the addition of the early stop callback option, thus avoiding possible overfitting. The batch size was determined as 80, which was obtained experimentally. Finally, to identify the strongest and weakest points of the classes, the confusion matrix was calculated.

## 4.2 Results

In figures 4 and 5, the grid search results are shown for the mel and MFCC descriptions, respectively.

The results indicate that the networks trained using the mel spectrogram exhibited exceptional performance on the training set when utilizing more than four convolutional layers. It is worth noting that the test set results revealed slight overfitting across all experiments. Nonetheless, the adaptive activation function, coupled with four convolutional layers and 400 filters, achieved an accuracy of 85.71%, which is a noteworthy achievement given the intricacy of the evaluated dataset.

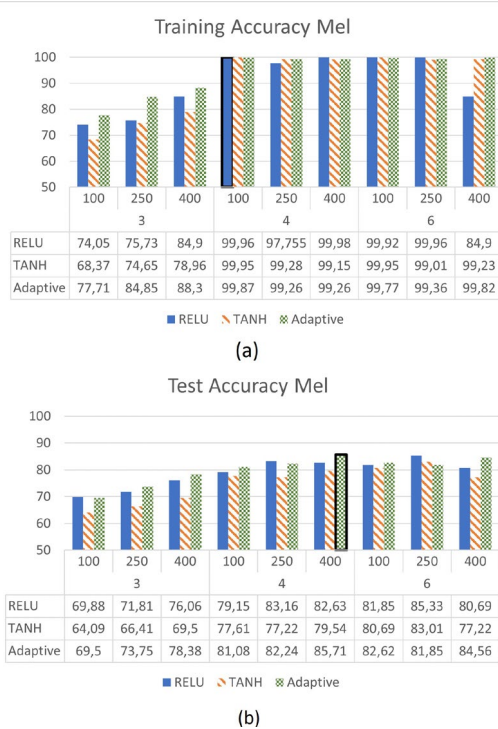

**Fig. 4** Training (a) and testing (b) accuracy for mel grid search of depths 3, 4 and 6; widths of 100, 250 and 400; and ReLU, TanH and an adaptive activation function.

Based on the experiments conducted, it was observed that fixing the depth of the network to four yielded the best results. Networks with three convolutional blocks did not perform as well, while increasing the number of blocks to six tended to result in overfitting. Additionally, increasing the number of filters was shown to improve network performance to a small extent in almost all cases.

Moreover, the adaptive activation function [36] was found to outperform classical methods in the classification task.

The loss and accuracy curves of the training are shown in Figure 6. The curves demonstrate the good performance of the training but are influenced by the dropout layer in all the peaks. Additionally, the plots show that the validation accuracy and loss follow a similar pattern to the training curves, indicating a slight overfitting on the training data. Overall, the loss and accuracy curves provide valuable insights into the performance of the model during training and can be used to optimize the hyperparameters of the network.

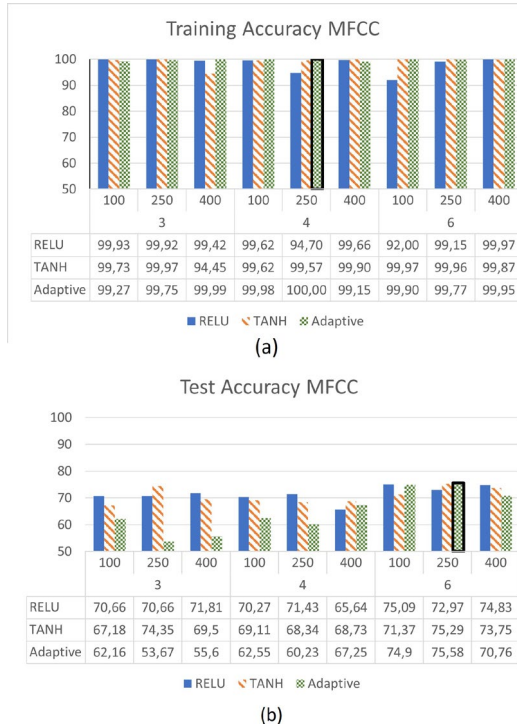

(a)

(b)

**Fig. 5** Training (a) and testing (b) accuracy for MFCC grid search for depths of 3, 4 and 6; widths of 100, 250 and 400; and ReLU, TanH and an adaptive activation function.

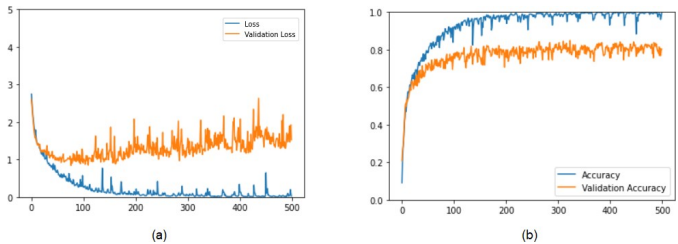

**Fig. 6** Loss and accuracy curves for training and validation (test) datasets for the network with the best hyperparameters

A comparative analysis was conducted between the mel spectrogram and MFCC features to identify the most effective feature extraction technique for the task at hand. The results are visually presented through a comparison chart (Figure 7). This study demonstrates that the mel spectrogram consistently outperformed all other approaches in all of the conducted experiments.

To determine the efficiency of the proposed method, the results were compared with those obtained by other known classifiers such as decision trees, naive Bayes, k-nearest neighbours (KNN), Gaussian mixture, random forest (RF), XGBoost, multilayer perceptrons (MLPs) and support vector classification (SVC). For all of the classical machine learning methods, the mel spectrogram description was used as the features of the audio.

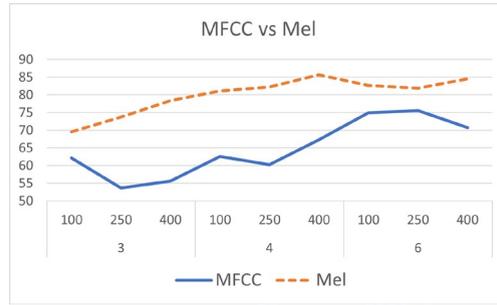

**Fig. 7** Mel spectrogram and MFCC results using the best activation function along different widths and depths.

In Figure 8, the results of the training and testing subsets are displayed. The bar represents the mean value of a 10-fold iteration. The standard deviation is displayed as the error bar. The model has some overfitting in both cases, which is more accentuated for the description with the MFCC. The best results were obtained for the mel spectrogram description, with 85.71% accuracy in the testing dataset.

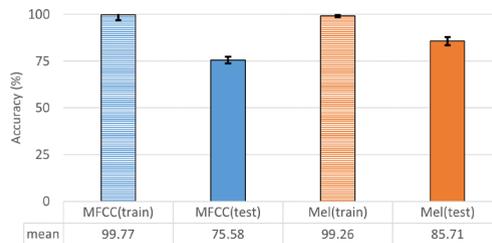

**Fig. 8** Training and testing accuracy for MFCC and mel spectrogram evaluation using the best hyperparameter configuration with the adaptive activation function.

In Figure 9, the accuracy obtained with all the classifiers and with our proposals is shown.

Comparing the proposed MFCC architecture (75.58%) with the best of the results obtained using classical methods (SVC with 51.03%), we can see that our method improves the result by 48.1%.

Now, taking into account the proposed mel spectrogram architecture (85.71%), which has a higher accuracy than the proposed MFCC architecture, the percentage of improvement of the architecture in this study with respect to the best result obtained with the classical methods is 67.96%, which represents considerable improvement in this area.

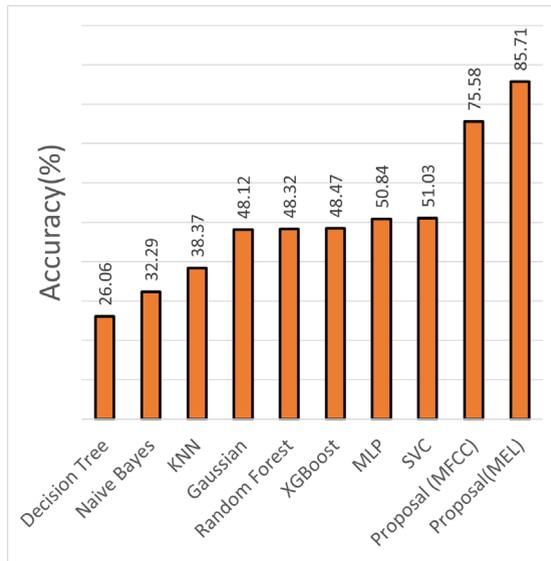

**Fig. 9** Results of bird classification using classical methods and our proposal based on the FCN.

In table 3, the results obtained in each of the 10 folds for the best hyperparameter configuration using mel and MFCC as descriptors are shown. As can be seen, there are iterations in which an accuracy percentage greater than 88% was obtained in the case of the mel spectrogram description, and in the case of MFCC, 79.00% was achieved.

**Table 3** 10-fold classification results using the mel spectrogram and MFCC as the descriptor

|      | 0     | 1     | 2     | 3     | 4     | 5     | 6     | 7     | 8     | 9     |
|------|-------|-------|-------|-------|-------|-------|-------|-------|-------|-------|
| Mel  | 86.45 | 87.75 | 82.85 | 83.00 | 85.56 | 88.09 | 86.12 | 82.31 | 87.27 | 87.65 |
| MFCC | 71.20 | 77.90 | 73.52 | 75.80 | 76.14 | 75.60 | 76.89 | 71.49 | 78.28 | 79.00 |

We calculated different metrics such as precision, recall, and F1-score for each of the test classes for the best model (see Table 4. The main purpose of this analysis was to identify the classes where the model makes the most mistakes and determine the potential areas of improvement.

To verify the good performance of the FCN, a neural network with the same convolutional layers but with fully connected layers as the classifier was tested.

The results are shown in Figure 10.

As shown in the results, the FCN outperformed the CNN in all experiments conducted.

The FCN allows predictions to be made with inputs of variable size. To verify the variation in the accuracy with respect to the duration of the audio files, a study was carried out using mel spectrograms on audio files of 1, 3, 5, 7, 10, 15 and 20 second durations. The results can be seen in Figure 11. It can be observed that the longer the audio files are, the better the results. However, the results improve in a fairly linear fashion, so the desired time can easily be determined from the penalty to be applied to the prediction system.

**Table 4** Precision, recall and f1-score for all the classes evaluated using the best hyperparameter configuration for the test set.

|  | precision | recall | f1-score | support |
|---|---|---|---|---|
| **Golondrina** | 0.67 | 0.80 | 0.73 | 5 |
| **Avion** | 0.91 | 1.00 | 0.95 | 10 |
| **Vencejo** | 1.00 | 0.82 | 0.90 | 17 |
| **Ciguena** | 0.94 | 0.94 | 0.94 | 18 |
| **CiguenaNegra** | 0.76 | 0.89 | 0.82 | 18 |
| **AguilaCalzada** | 0.78 | 0.74 | 0.76 | 19 |
| **Grulla** | 0.75 | 0.82 | 0.78 | 11 |
| **GaviotaReidora** | 1.00 | 1.00 | 1.00 | 6 |
| **Flamenco** | 0.00 | 0.00 | 0.00 | 1 |
| **Anade** | 0.94 | 0.83 | 0.88 | 18 |
| **Garza** | 0.67 | 0.80 | 0.73 | 15 |
| **Ansar** | 0.86 | 0.69 | 0.77 | 26 |
| **Correlimos** | 0.83 | 0.83 | 0.83 | 18 |
| **Cernicalo** | 0.88 | 0.79 | 0.83 | 19 |
| **Cuchara** | 0.75 | 0.82 | 0.78 | 22 |
| **Aguilucho** | 0.89 | 0.96 | 0.93 | 26 |
| **Andarríos** | 0.82 | 0.90 | 0.86 | 10 |
| **macro avg** | 0.79 | 0.80 | 0.79 | 259 |
| **weighted avg** | 0.84 | 0.85 | 0.85 | 259 |

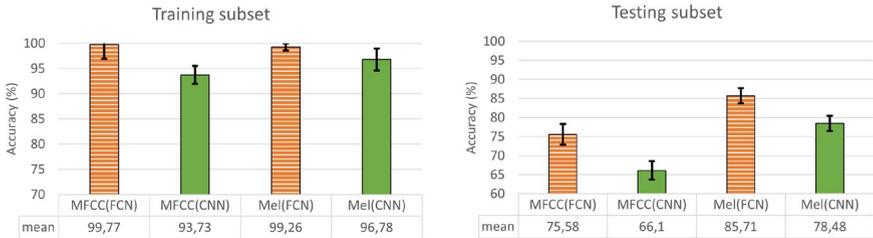

**Fig. 10** Results of bird classification using the FCN and CNN for both the mel and MFCC descriptions in the training and testing subsets.

## 4.3 Multispecies recognition

Fully convolutional neural networks have a major advantage in terms of the ability of the network to detect bird songs, even with different input lengths. This allows us to process a complete audio file to identify several species at the same time. Classic methods analyze the entire audio file to determine the most common bird by extracting audio features that are generally based on global features of the complete audio file. Our proposed method is able to identify not only the main bird in the audio, but also background birds or short songs of other species.

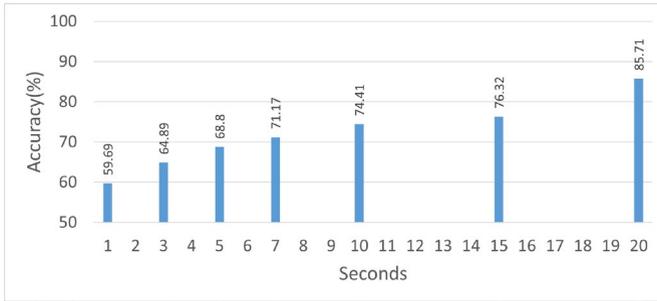

**Fig. 11** Accuracy in the test subset depending on the length of the audio file. Accuracies for 1, 3, 5, 7, 10, 15 and 20 seconds were obtained.

To evaluate the method proposed in this study, a concise classification of each test audio file was performed to determine all the birds in the entire file using the FCN over mel spectrogram images. Figure 12 shows an example of all the birds detected in an audio file labelled "Cigüeña" over time. In this example, two other species are detected, an "Ansar" and a short audio of a "Vencejo".

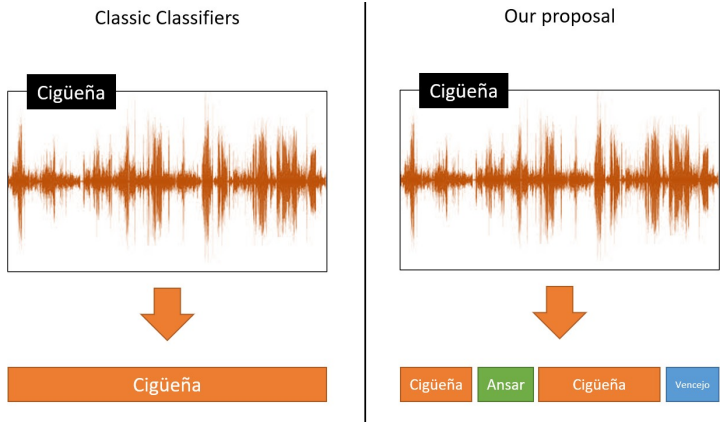

**Fig. 12** Real-time bird identification. With the method proposed in this study, the audio can be divided into chunks and classified independently.

In this experiment, the accuracy dropped to 49.73%, in contrast to the 82.16% achieved in the full audio classification with the mel spectrogram considering all the chunks of the audio belonging to the same class. This can be explained by the fact that our dataset was recorded in the wild. For this reason, an audio tagged "Ansar" may have other birds recorded in the background. Considering the good performance of our model with full audio files, this drop in the multiclass classification validates the correct performance of our model in identifying secondary bird songs.

In Figure 13, the confusion matrix of the multispecies recognition problem is shown.

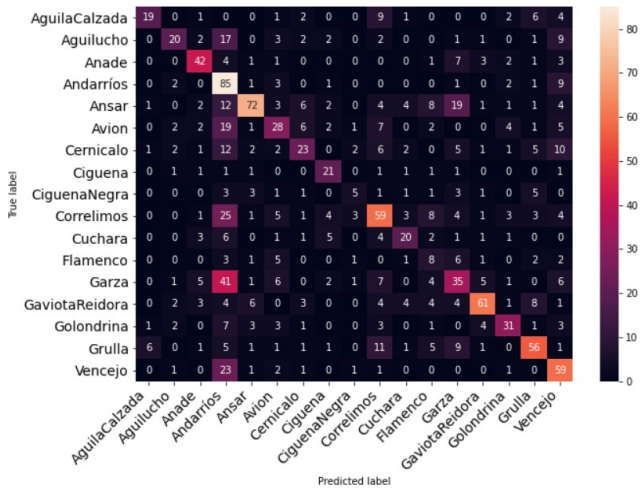

**Fig. 13** Confusion matrix of multispecies prediction.

The class "Andarrios" is usually detected as secondary in the audio files of other birds. Another example of this is the "Vencejo", which appears in almost every other class at least once.

In contrast, the "Aguila Calzada" and the "Aguilucho" seldom appear in other audio files. This is mainly due to the number of wild individuals in the area where the recordings were made and the behaviour of each species when making sounds.

Some of the failures were evaluated, and it was verified that in the cases in which the method failed, it was generally because the sound of another bird that the model correctly classified overlapped. This experiment was conducted with experts in the field. To verify this, a new prediction experiment using audio files made up of songs from different birds was performed. The results obtained in this experiment were similar to the results obtained with the original dataset (approximately 80% accuracy).

This method is very interesting since it is capable of detecting relationships or the appearance of different birds at the same time in an audio file. It is able to analyze more deeply the appearance of birds in a certain area without having to preprocess the recordings to separate the audio by type of sound and without the risk of a bird being hidden by a more relevant one in an audio file.

# 5 Conclusions

In this work, a new neural network architecture is proposed for bird identification based on song. The neural network used is a fully convolutional neural network, which allows input of variable length. This function allows us to not only feed the network audio of any length but also split any audio file into multiple short inputs to identify multiple species in the same audio file.

Mel spectrograms are used as descriptions of audio waves to allow the automatic extraction of powerful features to distinguish species thanks to the convolutional layers of the network. An extensive grid search evaluation was carried out regarding the best neural network configuration and the best hyperparameters. These tests determined that the best activation function for this problem is an adaptive activation function.

An exhaustive evaluation with classical state-of-the-art methods was carried out to determine the performance of our proposal, which achieved an improvement of more than 45% in all cases.

Furthermore, classical methods are only able to identify one species per audio file. The proposed method is able to detect multiple species' songs in the same audio file, improving the capability of the system to determine the presence of a specific bird without a complex preprocessing step for the audio.

The results are very interesting for the automatic identification of migratory birds taking into account only their songs.

In this work, a dataset with migratory species that pass through Spain was evaluated. In future works, we will add native species to the dataset to identify not only migratory birds but also the native birds of Spain.

In future works, the steps of this model will be carried out under a consistent approach in which preprocessing will be performed to separate the recorded audio according to silence thresholds that will allow us to identify continuous bird calls without the risk of analysing multiple birds in the same audio, except in the unavoidable case of overlaps.

Furthermore, the proposed model could be evaluated not only for this specific subset of birds but also for global bird classification. In addition, any type of echoacoustic classification task could be performed by training the appropriate audio due to the high adaptability of the proposed neural network.

# Declarations

## 5.1 Authors contribution statement

**María Teresa García-Ordás**: Conceptualization, Data curation, Methodology, Software, Visualization, Validation, Writing- Original draft preparation. **Sergio Rubio-Martín**: Data curation, Writing- Original draft preparation.**Héctor Alaiz-Moretón**: Conceptualization, Supervision, Writing-Reviewing and Editing. **Isaías García-Rodríguez**: Conceptualization, Supervision, Writing- Reviewing and Editing. **José Alberto Benítez-Andrades**: Data curation, Methodology, Software, Visualization, Validation, Writing- Reviewing and Editing.

## 5.2 Competing interests

The authors declare that they have no conflicts of interest regarding this work. The people involved in the experiments were informed and formally consented.

## 5.3 Availability of data and materials

The code and data are available from the corresponding author on reasonable request.

## 5.4 Ethical and informed consent for data used

Not applicable.